# Environment-assisted crack nucleation in La(Fe,Mn,Si)$_{13}$-based magnetocaloric materials

Siyang Wang[1,*], Edmund Lovell[2], Liya Guo[3], Neil Wilson[2], Mary P. Ryan[1], Finn Giuliani[1]

[1] Department of Materials, Royal School of Mines, Imperial College London, London, SW7 2AZ, UK

[2] Camfridge Ltd., Unit B1, Copley Hill Business Park, Cambridge Rd, Babraham, Cambridge, CB22 3GN, UK

[3] School of Materials Science and Engineering, Shanghai University, Shanghai 200444, China

*Corresponding author: siyang.wang15@imperial.ac.uk

# Abstract

Cracking in La(Fe,Si)$_{13}$-based magnetocaloric materials has been observed to predominantly form around La-rich (La$_2$O$_3$) particles and pose a threat to their long-term structural integrity. To understand the formation of these cracks, FIB cross-sectional polishing followed by SEM characterisation was employed to study local microstructural evolution after air exposure. Results suggest that volume expansion and internal degradation associated with a chemical reaction between La$_2$O$_3$ particles and water/moisture can lead to crack nucleation in the 1:13 phase adjacent to La-rich particles. This observation indicates that the formation of La-rich phase should be suppressed, or their size minimised during material processing to ensure the long-term structural integrity of La(Fe,Mn,Si)$_{13}$ magnetocaloric materials.

Keywords: Magnetocalorics; Materials; Environment; Crack; Mechanical stability

Magnetic refrigeration is a promising next-generation cooling technology which avoids the use of fluorinated gases, and its potential for higher energy efficiency when compared to conventional gas compression-based technology is promising with regard to lowering CO$_2$ emission *via* reduced electricity consumption (Balli et al., 2017; Franco et al., 2018; Gschneidner and Pecharsky, 2008). As a candidate material system for magnetic refrigeration, La(Fe,Mn,Si)$_{13}$ intermetallics have advantages of low cost, a sharp first-order magnetic phase transition, a giant magnetocaloric effect, and a tuneable operating temperature range near room temperature (Bratko et al., 2017; Fujita et al., 2003; Guo et al., 2020; Liu et al., 2012).

Nevertheless, these materials are mechanically brittle, and therefore their structural integrity over long-term servicing is a major concern. During manufacturing and operational processes, potential origins for crack development include mechanical load upon machining, thermal stress during hydrogenation (a necessary step to bring the operating temperature close to room temperature), and volume change incompatibility at the magnetic phase transition (Kitanovski, 2020; Lyubina et al., 2010). However, it was found that there are already pre-existing cracks around La-rich phase in the raw material before final shaping (Glushko et al., 2019). Such cracks could act as initiators for catastrophic failure, therefore their elimination from the microstructure would be desirable. Whilst they have been observed, their formation mechanism remains unknown. Specifically, as these materials are immersed in heat exchange fluid during operation of magnetocaloric devices, it is important to understand whether the formation of those cracks are related to interactions between the material and the environment.

La(Fe,Mn,Si)$_{13}$-based magnetocaloric materials were received from Camfridge Ltd. Hydrogenation was performed by Vacuumschmelze GmbH & Co. KG *via* heat treatment in hydrogen atmosphere (reported in (Barcza et al., 2011)). The temperature where the adiabatic temperature change for a 0 to 1.5 T field change is maximum ($T_{peak}$) for the hydrogenated material studied here is ~15.6 °C. All the sample preparation, storage and examination steps were carried out at room temperatures of 21 ± 3 °C, thus the samples employed should not have undergone the magnetic phase transition since they were received.

Specimens were polished under vacuum using focussed ion beam (FIB) cross-sectioning, on a FEI Helios Nanolab 600 Dualbeam microscope. FIB cross-sectioning was employed to avoid exposure of the polished surface to air and water. The samples were separated into two groups: one group of samples were taken out from the microscope when the polishing was complete and stored in air, while the others were left in the microscope without venting the chamber and were hence stored under vacuum (~$5.25e^{-5}$ Pa). Energy-dispersive X-ray spectroscopy (EDX) was used to locate the phases present, on a Zeiss Sigma 300 scanning electron microscope (SEM) equipped with a Bruker XFlash 6-60 EDX detector. Microstructural evolution in the FIB-polished areas over time was captured through secondary electron imaging, using either the Sigma SEM for the samples stored in air, or directly in the Helios Dualbeam microscope for the samples remained in vacuum.

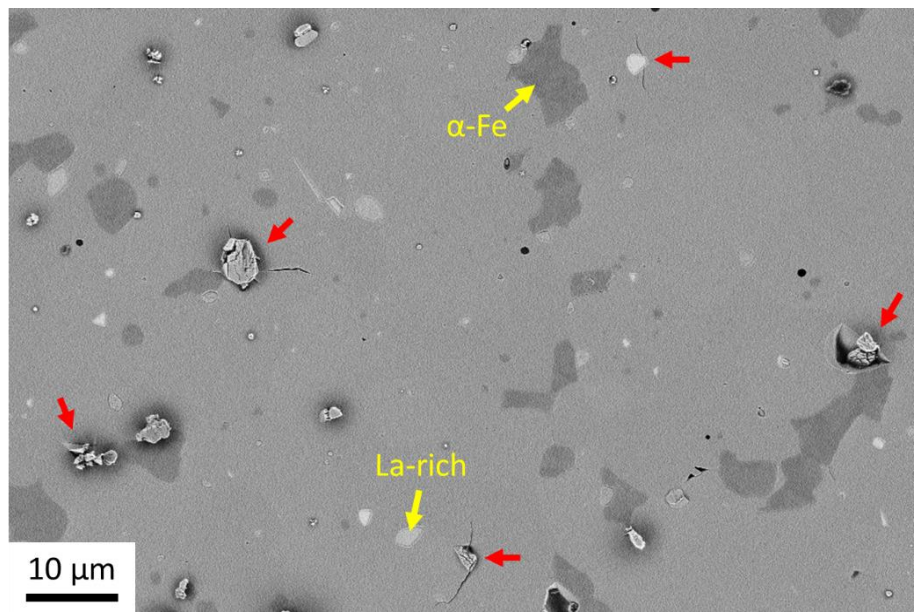

Figure 1 Backscattered electron (BSE) image showing the microstucture of the hydrogenated material, after 1-hour-polishing with colloidal silica. The 1:13, α-Fe, and La-rich phases correspond to the medium, low, and high brightness areas, respectively. Cracks within and around La-rich particles can be observed, as marked with the red arrows.

The microstructure of the specimens (hydrogenated, top surface polished with colloidal silica) is shown in Figure 1. Three phases can be observed: main (1:13) phase (medium brightness), α-Fe phase (low brightness), and La-rich phase (high brightness). Cracks within and around

La-rich particles are clearly visible, as marked with the red arrows (consistent with work reported in (Glushko et al., 2019)).

The evolution of freshly FIB-polished cross-sectional microstructure in response to air exposure, as characterised by a series of SEM images, is given in Figure 2 (at 1- and 7-days post-FIB-cross-sectional-polishing). EDX map for La distribution shows the locations of the phases: high-, medium- and low-La content areas correspond to La-rich, 1:13, and α-Fe phases, respectively. After 1-day exposure to air, two La-rich particles, as marked with the black arrows, protruded from the surface and cracks were visible around the particles. 6 days later, the internal cracks (cracks *within* the La-rich particles) and the extent of the protrusion increased. Moreover, these dynamic behaviours have resulted in the formation of cracks within the 1:13 phase around the La-rich particles, as marked with the red and yellow arrows on the image. Cracking near the top edge of the sample (yellow arrow) was significant enough to cause the local region to be broken and pushed upwards.

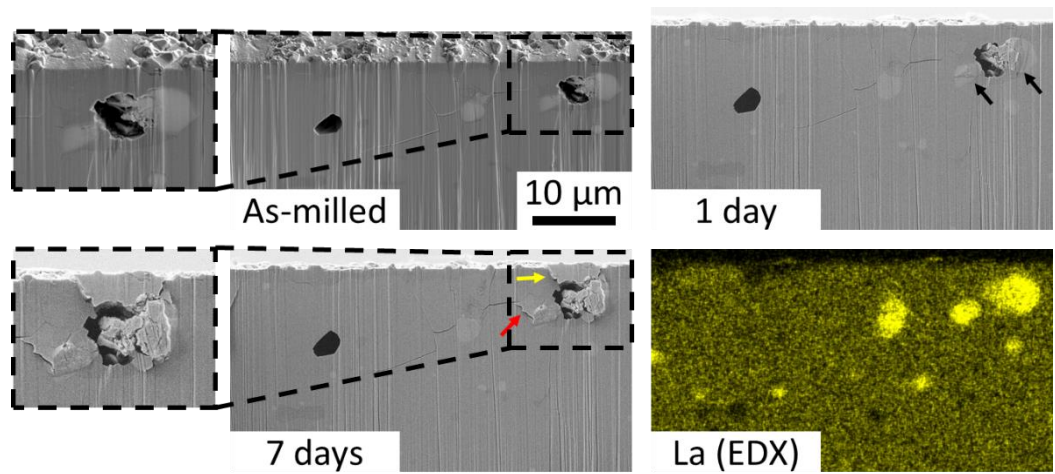

Figure 2 Microstructural evolution in a FIB-polished area on a hydrogenated sample during air exposure, characterised with SEM imaging (case 1). EDX map of La distribution reveals the locations of the phases.

Figure 3 shows another example. The La-rich particle marked with the red arrow on the "day 1" image protruded from the surface when exposed to air, which gradually became more significant over time. This later led to a fraction of the 1:13 phase on its right being torn apart, as well as the formation of fringes in the α-Fe phase on the top, as marked with the white and green arrows on the "6 days" image, respectively. Additionally, it is observed that not all La-rich particles that showed this behaviour started to exhibit internal crack and protrude from the surface shortly after FIB polishing, but early evidence of those might not exist until the sample was exposed to air for a few days, as marked with the yellow arrows on the images.

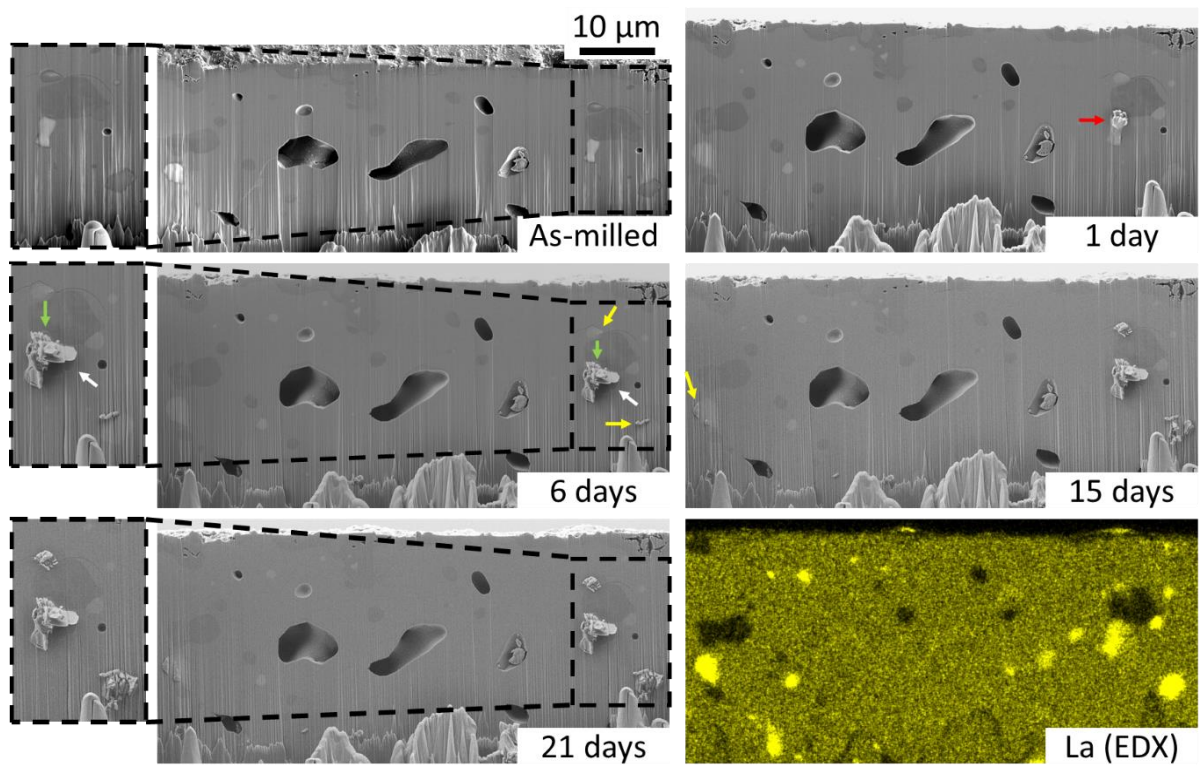

Figure 3  Microstructural evolution in a FIB-polished area on a hydrogenated sample during air exposure, characterised with SEM imaging (case 2). EDX map of La distribution reveals the locations of the phases.

For the other group of samples which remained under vacuum, the phenomena shown above was not observed up to 60 h post-polishing. Figure 4 shows an example of this, where the surface remained flat and no evidence of crack development within/around La-rich particles or any other microstructural features can be seen in the first 3 images. Some pre-existing cracks can be observed within the La-rich particles near the pores, as marked with the white arrows on the "as-milled" image. The sample was then taken out from the microscope, stored in air for another 60 h, and characterised again. It can be seen from the bottom right image in Figure 4 that after air exposure for 60 h, a few La-rich particles protruded from the surface, including but not limited to the ones marked with the yellow arrows. The protrusion of some of the larger particles had initiated damage in the neighbouring 1:13 matrix phase, such as the one marked with the red arrow which caused the cracking of the 1:13 phase beneath it. Another general observation for all the results shown above, is that most of the La-rich particles did not have internal cracks in the as-FIB-polished state, while the ones that are originally defective are *all* adjacent to or inside pores.

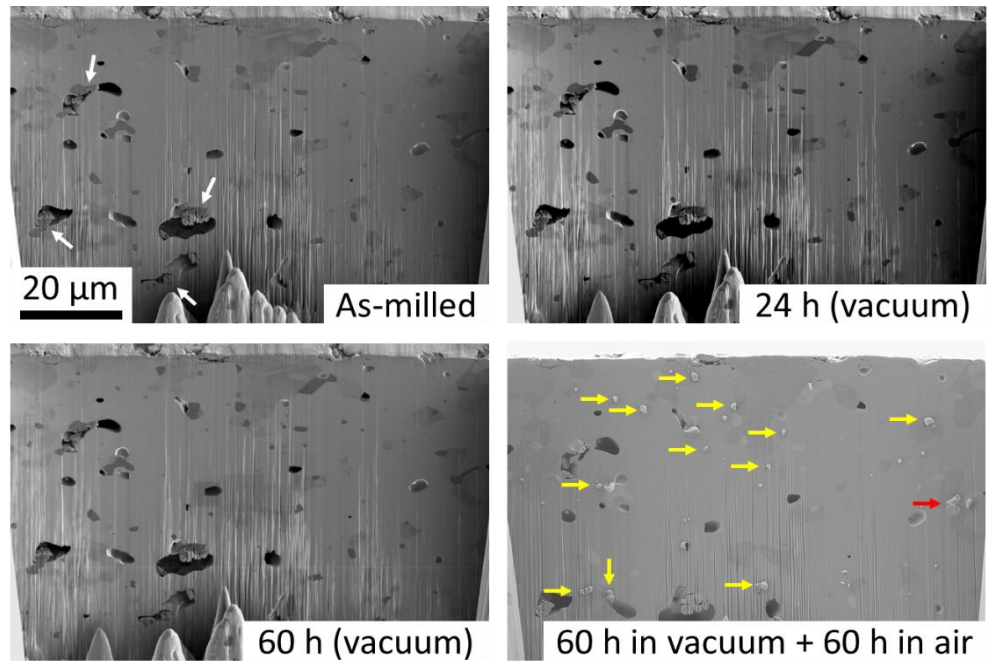

Figure 4 Microstructural evolution in a FIB-polished area for a hydrogenated sample stored in the dualbeam microscope under vacuum (~5.25e$^{-5}$ Pa) for 60 h and then in air for another 60 h, characterised with SEM imaging.

The nature of the La-rich phase is known to be $La_2O_3$, a hygroscopic compound that easily picks up moisture in the air and transforms into $La(OH)_3$ (Löwe et al., 2012). This is consistent with our results where this reaction is supressed when air/moisture source is removed. However, through controlling the environment of the sample preparation steps, we have demonstrated that the "pre-existing" cracks within and around La-rich particles observed after mechanical polishing (Glushko et al., 2019) are not real reflections of the material microstructure, but likely environment-induced artefacts owing to the volume expansion and internal degradation associated with the hydroxylation reaction. It is evident that the significant volume expansion (~72% according to (Koehler and Wollan, 1953; Zachariasen, 1948)) caused severe damage in the surrounding 1:13 phase, which, during subsequent processing and operation that subject the material to local stresses, could lead to crack propagation and even catastrophic failure such as those reported recently in (Lionte et al., 2021). This phenomenon of crack development within and around $La_2O_3$ particles is not restricted to the ones exposed on the surface, but also occurs for those near the pores *inside* the material although in this scenario the local moisture source should be finite, and it is not clear yet if there is a different mechanism for crack nucleation in this instance. It is anticipated that during operation of magnetic refrigeration devices where magnetocaloric materials are immersed in water-based heat transfer fluid, this process of crack nucleation would be accelerated due to the much more abundant $H_2O$ as compared to moisture in air. Moreover, since the materials employed in magnetocaloric regenerators are often in the form of regular particles (Liang et al., 2021) or very thin plates (Engelbrecht et al., 2011) in order to access high energy efficiency, for the latter these cracks formed on surfaces and around pores can have pronounced effects on the overall mechanical stability. However, we also observed that crack nucleation within the main phase is less-frequently observed around small (sub-micron-sized) $La_2O_3$ particles, and therefore this detrimental effect of La-rich particles may be minimised through controlling their size range *via* refinement of the treatment parameters.

In summary, through FIB cross-sectional polishing under vacuum and subsequent microstructural characterisation with SEM, we found that the observed cracks within and around La-rich particles in La(Fe,Si)$_{13}$-based magnetocaloric materials are not intrinsic features of the original microstructure, but originate from exposure of the material to air/moisture. Volume expansion and internal degradation associated with the $La_2O_3$-$H_2O$ reaction are significant enough to induce crack nucleation events in the surrounding 1:13 phase which may act as precursors to catastrophic failure. Formation of the La-rich phase should be suppressed, or their size minimised during material production to ensure long-term structural integrity.

## Acknowledgements

The authors acknowledge funding from Innovate UK (UKRI 32645).